\documentclass{pasj01}
\Received{\today}
\Accepted{$\langle$acception date$\rangle$}
\Published{$\langle$publication date$\rangle$}
\begin{document}

\title{Optimum Frequency of Faraday Tomography to Explore the Inter-Galactic Magnetic Field in Filaments of Galaxies}
\author{Takuya \textsc{Akahori}\altaffilmark{1}\thanks{Corresponding author: takuya.akahori@nao.ac.jp}, Shinsuke \textsc{Ideguchi}\altaffilmark{2,3}, Takahiro \textsc{Aoki}\altaffilmark{4}, Kazuhiro \textsc{Takefuji}\altaffilmark{5}, Hideki \textsc{Ujihara}\altaffilmark{5}, and Keitaro \textsc{Takahashi}\altaffilmark{3}
}%
\altaffiltext{}{$^{1}$Mizusawa VLBI Observatory, National Astronomical Observatory of Japan, 2-21-1 Osawa, Mitaka, Tokyo 181-8588, Japan \\
$^{2}$Center for Computational Astrophysics, National Astronomical Observatory of Japan, 2-21-1 Osawa, Mitaka, Tokyo 181-8588, Japan \\
$^{3}$Department of Physics, Kumamoto University, 2-39-1, Kurokami, Kumamoto 860-8555, Japan\\
$^{4}$The Research Institute for Time Studies, Yamaguchi University, 1677-1, Yoshida, Yamaguchi, Yamaguchi 753-8511, Japan\\
$^{5}$Kashima Space Technology Center, National Institute of Information and Communications Technology, 893-1, Hirai, Kashima, Ibaraki 314-8501, Japan\\
}
%%\email{takuya.akahori@nao.ac.jp}

\maketitle

%%%%%%%%%%%%%%%%%%%%%%%%%%%%%%%%%%%
\begin{abstract}
Faraday tomography is thought to be a powerful tool to explore cosmic magnetic field. Broadband radio polarimetric data is essential to ensure the quality of Faraday tomography, but such data is not easy to obtain because of radio frequency interferences (RFIs). In this paper, we investigate optimum frequency coverage of Faraday tomography so as to explore Faraday rotation measure (RM) due to the intergalactic magnetic field (IGMF) in filaments of galaxies. We adopt a simple model of the IGMF and estimate confidence intervals of the model parameters using the Fisher information matrix. We find that meaningful constraints for RM due to the IGMF are available with data at multiple narrowbands which are scattered over the ultra-high frequency (UHF, 300~MHz -- 3000~MHz). The optimum frequency depends on the Faraday thickness of the Milky Way foreground. These results are obtained for a wide brightness range of the background source including fast radio bursts (FRBs). We discuss the relation between the polarized-intensity spectrum and the optimum frequency.
\end{abstract}

%%%%%%%%%%%%%%%%%%%%%%%%%%%%%%%%%%%
%%%%%%%%%%%%%%%%%%%%%%%%%%%%%%%%%%%
\section{Introduction}
\label{s1}

\begin{figure*}[t]
\begin{center}
\includegraphics[width=160mm]{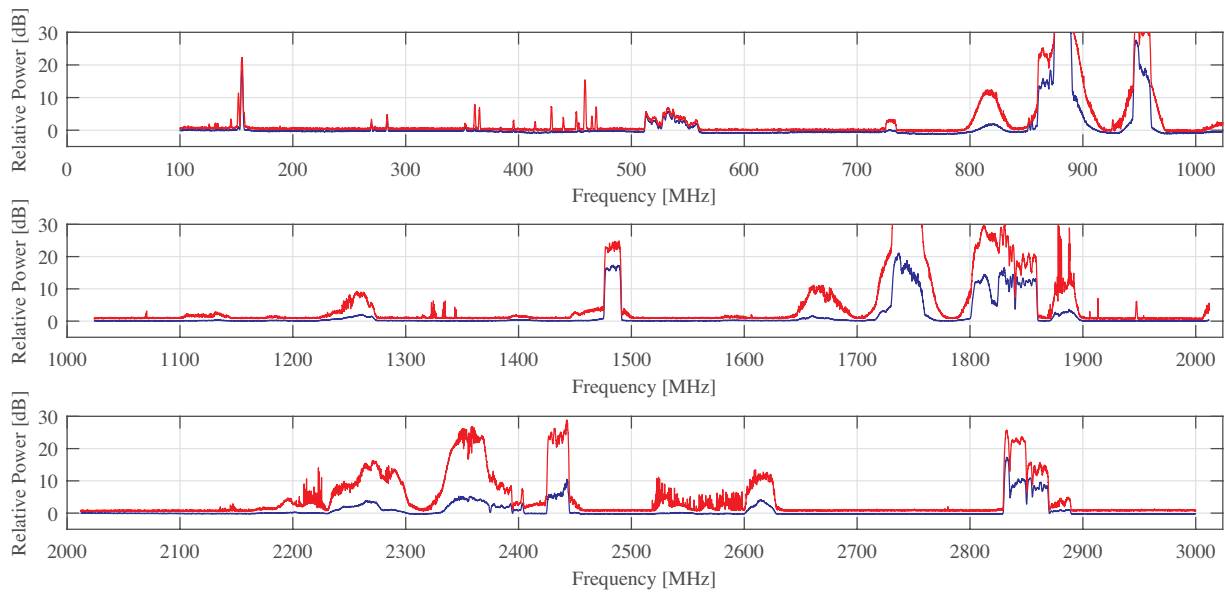}
\end{center}
\caption{
Radio frequency environment around the Kashima 34-m antenna of the National Institute of Information and Communications Technology (NICT) in Japan. The blue and red lines show the instantaneous (sweep time 2.05 second) and 5 minutes max-hold (i.e. maximum during 5 minutes) spectra, respectively. The band characteristic of the receiver is removed from the spectra, so that the vertical axis is the relative radio power with respect to the detection limit.
}
\label{f01}
\end{figure*}

Magnetic field is a fundamental element of the Universe and it affects formation and evolution of astronomical objects. Centimeter radio polarimetry is one of the promising tools to study cosmic magnetism (see \cite{han17, aka18} for reviews); synchrotron intensity, its linear-polarization vector, and Faraday rotation measure (RM) provide us with properties of magnetic field in galaxies and AGN jets, and they reveal detailed structures of magnetized plasma such as the interstellar medium (ISM) and intergalactic medium (IGM). Cosmic magnetism is one of the key sciences for the Square Kilometre Array (SKA) \citep{joh15}.

Faraday RM synthesis or Faraday tomography (\cite{bur66, bre05}) grows up progressively in radio polarimetry. There are a lot of successful applications to the ISM \citep{sak18}, galaxies \citep{mao17}, radio lobes \citep{osu18}, quasars \citep{and15, and16}, and galaxy clusters \citep{oza15}. Furthermore, discovery of Fast radio bursts (FRBs) fosters momentum of the study of cosmic magnetism. As at April 2018, seven linearly-polarized FRBs are published in the literature (see \cite{cal18}). For example, \citet{mic18} observed FRB121102 and found strongly-magnetized medium with ${\rm RM} \sim O(10^5)$~${\rm rad~m^{-2}}$, implying an environment similar to that around a super massive black hole (SMBH).

It has been predicted that the cosmic web is permeated with a large amount of magnetized IGM. \citet{aka14a} studied possible situations to estimate RM due to the intergalactic magnetic field (IGMF) by means of Faraday tomography, and demonstrated that the ultra-high frequency (UHF) band is promising to maximize the capability of Faraday tomography for the study. \citet{rav16} applied Faraday tomography to FRB150807 and derived an upper limit of the IGMF strength $<21$~nG, which does not contradict theoretical predictions (e.g., \cite{aka16, vaz18}).

The above studies demonstrate the capability of Faraday tomography for a wide RM range of diffuse, compact, and even time-domain radio sources. Because wider frequency coverage gives better quality of Faraday tomography (e.g., \cite{aka14a}), a modern wideband observation makes Faraday tomography feasible. However, obtaining a seamless dataset over broad bandwidth is difficult. One of the essential reasons is radio frequency interferences (RFIs). Centimeter wavelength is commonly used in industry, such as broadcasting, mobile phone, wireless communication, and radar. Figure~\ref{f01} shows an example of RFIs (see Appendix for observational details). Appreciable signals are all RFIs against radio astronomy. These RFIs easily saturate amplifiers, produce artificial higher-harmonic signals, and alter the band characteristics fatally; they make signal processing unreliable.

Persistent RFIs can be cut by frequency filters at an early stage of a receiver system, but this means that we never obtain astronomical signal at the frequencies. Although many large radio telescopes are located at countryside with low human population, radio frequency environment rapidly changes as human lifestyle improves\footnote{Indeed, ``Sky Muster" RFIs at ATCA 15 mm band and ``BSAT-4a" RFIs at VERA K band are very recently appeared.}. SKA-MID antennas will be constructed in radio-quiet districts in South Africa, but economic growth in South Africa would impact on radio frequency environment at the site in future. 

In this paper, we investigate the optimum frequency of Faraday tomography to explore RM due to the IGMF. Although \citet{aka14a} briefly considered RFIs on the SKA sites, a more comprehensive study about frequency dependence on Faraday tomography could maximize the chance to discover the IGM and IGMF through Faraday tomography. This paper is organized as follows. We explain our model and calculation in Section 2. The results are shown in Section 3 followed by discussion and summary in Section 4.

%%%%%%%%%%%%%%%%%%%%%%%%%%%%%%%%%%%
%%%%%%%%%%%%%%%%%%%%%%%%%%%%%%%%%%%
\section{Model and Calculation}
\label{s2}

Our model and calculation are mostly the same as the previous works \citep{aka14a, ide14a}. \citet{aka14a} studied two cases of observations, (i) background compact source behind diffuse foreground source, and (ii) two pair compact sources. This paper addresses the case (i). Below, we briefly review our model and calculation. Readers who want to know more details can read the above references. 

%%%%%%%%%%%%%%%%%%%%%%%%%%%%%%%%%%%
\subsection{Model}
\label{s2.1}

We construct our model in the domain of Faraday dispersion function (FDF) or Faraday spectrum, $F(\phi)$. The FDF in units of ${\rm Jy~rad^{-1}~m^{2}}$ is defined in the form of Fourier transform,
\begin{equation}\label{eq01}
P(\lambda^2) 
=\int_{-\infty}^{\infty} F(\phi) e^{2i\phi \lambda^2} d\phi,
\end{equation}
where $P(\lambda^2) = Q(\lambda^2)+iU(\lambda^2)$ is the complex polarized intensity of the Stokes parameters $Q$ and $U$ in Jy, and $\lambda$ is the wavelength in meter. The Faraday depth (RM), $\phi(x)$, in ${\rm rad~m^{-2}}$ is defined as
\begin{equation}\label{eq02}
\phi(x)=810 \int_{x}^{0}n_{\rm e}(x')B_{||}(x')dx',
\end{equation}
where $n_{\rm e}$ is the thermal electron density in ${\rm cm^{-3}}$, $B_{||}$ is the line-of-sight (LOS) magnetic field strength in ${\rm \mu G}$, 
and $x'$ is the LOS physical distance in kpc.

\begin{figure}[tp]
\begin{center}
\FigureFile(80mm,40mm){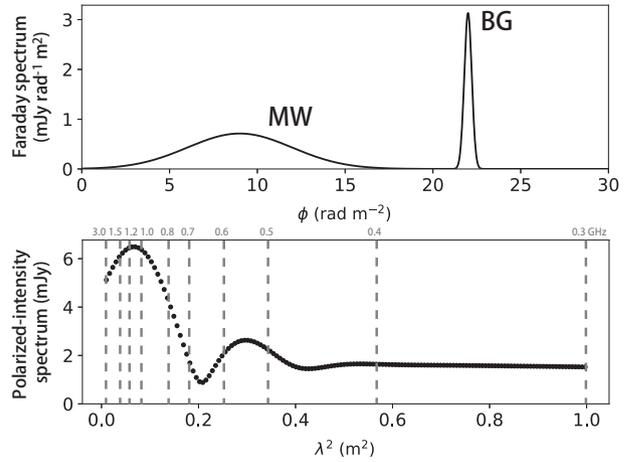}
\end{center}
\caption{
An example of a Faraday spectrum (top) and a polarized-intensity spectrum (bottom).
}
\label{f02}
\end{figure}

The top panel of figure~\ref{f02} shows an example of an FDF model. We assume to select spatially-compact sources such as quasars, radio galaxies, or FRBs and choose an ideal Faraday-thin source whose thickness in the Faraday-depth space is sufficiently small compared to the resolution of the Faraday depth (see, e.g. \cite{aka18}). We also suppose that there is no intervening galaxy along the LOS toward the source. This source then appears at a certain Faraday depth induced by RMs due mostly to the IGMF and the Milky Way. The Milky Way can be bright and likely appear as a Faraday-thick source, so that the gap between the two signals in $\phi$ space becomes a measure of RM due to the IGMF (\cite{aka14a}; see also \cite{aka14b} for source selection strategy).

The above situation is modeled with two Gaussian functions as follows:
\begin{equation}\label{eq03}
F(\phi) = \sum_{i={\rm MW,BG}} \frac{f_i e^{2 i \theta_i}}{\sqrt{2\pi} \delta\phi_i}\exp\left\{ -\frac{(\phi-\phi'_i)^2}{2\delta\phi_i^2} \right\},
\end{equation}
where the subscripts MW and BG represent a Milky Way foreground and an extragalactic background, respectively. Each source is characterized by the Faraday depth up to the center of the source ($\phi'$ in ${\rm rad~m^{-2}}$), the Faraday thickness of the source ($\delta \phi$ in ${\rm rad~m^{-2}}$), the amplitude ($f$ in ${\rm Jy~rad^{-1}~m^{2}}$), and the intrinsic polarization angle ($\theta$ in radian). Therefore, our FDF model consists of total eight parameters. More complicated models will be briefly discussed in Section 4.

RM due to the IGMF is defined by truncating the Gaussian tail at three times of the standard deviation:
\begin{equation}\label{eq04}
RM_{\rm IGMF}
= \left( \phi'_{\rm BG} - 3 \delta \phi_{\rm BG} \right)
  - \left( \phi'_{\rm MW} + 3 \delta \phi_{\rm MW} \right),
\end{equation}
We consider $RM_{\rm IGMF}$ from 1 ${\rm rad~m^{-2}}$ to 20 ${\rm rad~m^{-2}}$ covering cases for a single galaxy filament \citep{aka10} and multiple filaments toward high redshift \citep{aka11}. The choice of $\phi'_{\rm MW}$ and $\delta \phi_{\rm MW}$ correspond to a choice of sky position; the Milky Way contribution toward the south Galactic pole is $\phi'_{\rm MW} \sim 6$~${\rm rad~m^{-2}}$ and $3 \delta \phi_{\rm MW} \sim 5$~${\rm rad~m^{-2}}$ (\cite{aka13, opp15}) and those are larger at lower galactic latitudes in general. We consider $\delta \phi_{\rm MW}$ from 2 ${\rm rad~m^{-2}}$ to 12 ${\rm rad~m^{-2}}$. The model shown in figure~\ref{f02} is an example in these parameter ranges; $\phi'_{\rm MW}$ = 9 ${\rm rad~m^{-2}}$, $\delta \phi_{\rm MW}$ = 7 ${\rm rad~m^{-2}}$, $\phi'_{\rm BG}$ = 22 ${\rm rad~m^{-2}}$, and $\delta \phi_{\rm BG}$ = 0.5 ${\rm rad~m^{-2}}$.

%%%%%%%%%%%%%%%%%%%%%%%%%%%%%%%%%%%
\subsection{Calculation}
\label{s2.2}

We adopt a model-fitting method in which we compare the observed data with a numerical model and find the best model parameters that minimize the Chi-squared value. In reality, the transformation from the observed Stokes Q and U to the FDF is always imperfect due to incompleteness of wavelength coverage, while the transformation from a model FDF to Stokes Q and U can reduce this numerical error because of less incompleteness of Faraday-depth coverage. Therefore, we compare the observed (mock) polarized-intensity spectrum with the spectrum given by our model FDF. This approach is called ``QU-fitting".

The bottom panel of Fig.~\ref{f02} shows an example of the polarized-intensity spectrum derived from a Fourier transform of the example FDF (figure~\ref{f02} top). To this synthesized spectrum, we add two observational effects, frequency coverage and noise, and construct a mock polarization spectrum data, as follows.

\begin{table}
\tbl{The frequency band definition in this work.}{%
\begin{tabular}{l|cccc}
\hline
\hline
Band & minimum & center & maximum & bandwidth\\
& MHz & MHz & MHz & MHz\\
\hline
P$_*$ & 300 & --- & 1000 & 20 or 40 \\
L & 1000 & 1500 & 2000 & 1000 \\
L$_{\rm 14}$ & 1400 & 1410 & 1420 & 20 \\
L$_{\rm 16}$ & 1500 & 1550 & 1600 & 100 \\
S$_{\rm 27}$ & 2650 & 2700 & 2750 & 100 \\
\hline
\end{tabular}}\label{t01}
\end{table}

Frequency coverage is the main concern of this work. According to the previous work \citep{aka14a}, we consider the UHF (300~MHz -- 3000~MHz) band which is promising for the study of the IGMF. Particularly, we examine the optimum frequency at low frequency (300~MHz -- 1000~MHz) in the UHF, because the dependence of the polarized-intensity spectrum on $\lambda^2$ is significant in this band (see Fig.~\ref{f02} bottom). Meanwhile, at high frequency (1000~MHz -- 3000~MHz), the dependence is minor and fine-tuning of the band selection does not significantly change the result. Therefore, we employ fixed, given narrowbands which are motivated by radio quietness\footnote{Part of them are recognized for usage of radio astronomy as primary (passive) or secondary.} at high frequency in the UHF (e.g., Fig.~\ref{f01}). Table~\ref{t01} summarizes the band definition in this work. We consider one broadband (L) and/or three narrowbands labelled L$_{\rm 14}$, L$_{\rm 16}$, and S$_{\rm 27}$. Here, throughout this work, one frequency channel or the frequency resolution is 1~MHz following modern large polarization surveys (e.g., polarization sky survey of the Universe's magnetism, POSSUM). We then explore the best center frequency of the P$_*$ band with a narrow bandwidth of 20~MHz or 40~MHz (20 channels or 40 channels). The above frequency coverage is applied to the Faraday spectrum by convolution using the window function approach \citep{aka14a}. 

Observational noise, or the sensitivity, is considered as follows. We consider flat frequency spectra of both MW and BG sources, meaning that the intrinsic polarized-intensity of the sources does not change in frequency. We add a random gaussian noise, where the signal-to-noise ratio in each 1 MHz channel is 10 for MW and 100, 1000, or 10000 for BG. Hence, if we suppose a 100~$\mu$Jy noise level, we are considering the Milky Way foreground of 1~mJy and the background source of from 10~mJy to 1~Jy.

Finally, we attempt to evaluate error profiles of model parameters using the Fisher information matrix. That is, for $\vec{p}=(p_1, p_2, \cdots, p_i, \cdots, p_j, \cdots)$ as a set of model parameters, the covariance between the $i$-th and $j$-th parameters is given by $\sigma_{ij}^2=\left(\mathcal F^{-1}\right)_{ij}$ and the marginalized 1-$\sigma$ error of the $i$-th parameter ($j=i$) is $\sigma_{ii}$. The Fisher information matrix is written as 
\begin{eqnarray}\label{eq05}
\mathcal F_{ij}
=\sum_{l=1}^N
\left[
\frac{1}{{\sigma^2(\lambda^2_l)}}
\left\{
\frac{\partial Q(\lambda^2_l;\vec{p})}{\partial p_i}\frac{\partial Q(\lambda^2_l;\vec{p})}{\partial p_j} \right. \right. \nonumber \\
\left. \left. + \frac{\partial U(\lambda^2_l;\vec{p})}{\partial p_i}\frac{\partial U(\lambda^2_l;\vec{p})}{\partial p_j}
\right\}_{\vec{p} = {\hat p}}
\right],
\end{eqnarray}
where $N$ is the number of $\lambda^2$ channels corresponding to the number of frequency channels we considered. A Gaussian likelihood is assumed and the gradients are calculated at the fiducial set of parameters, ${\hat p}$, around which confidence intervals are put \citep{ide14a}. Throughout this paper, we adopt 3-$\sigma$ confidence intervals of model parameters, i.e. $3\sigma_{ii}$. Error propagation to $RM_{\rm IGMF}$ is calculated from equation~(\ref{eq04}) with $3\sigma_{ii}$ errors of the relavant model parameters.

%%%%%%%%%%%%%%%%%%%%%%%%%%%%%%%%%%%
%%%%%%%%%%%%%%%%%%%%%%%%%%%%%%%%%%%
\section{Result}
\label{s3}

%%%%%%%%%%%%%%%%%%%%%%%%%%%%%%%%%%%
\subsection{Optimum Frequency of P$_*$ band}
\label{s3.1}

\begin{figure*}[t]
\begin{center}
\FigureFile(170mm,170mm){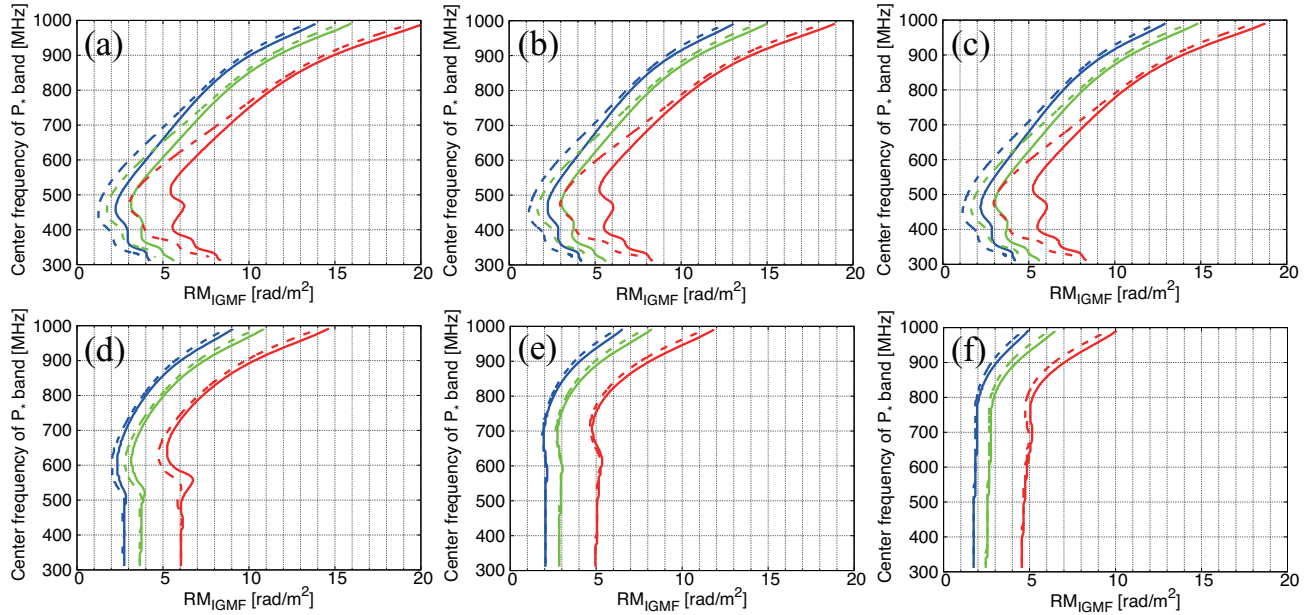}
\end{center}
\caption{
Error profiles between the input $RM_{\rm IGMF}$ and the chosen center frequency of P$_*$ band. Results in the cases of the P$_*$ + L bands are shown. The blue, green, and red lines show the contours on which $RM_{\rm IGMF}$ is determined with statistical errors of 30~\%, 20~\%, and 10~\%, respectively. The solid and dashed lines are the results with 20~MHz and 40~MHz bandwidths of P$_*$ band, respectively. Panels (a), (b), and (c) are the results for $F_{\rm BG}/F_{\rm MW} =$ 10, 100, and 1000, respectively, with $\delta\phi_{\rm MW}=$ 2 ${\rm rad~m^{-2}}$. Panels (d), (e), and (f) are the results for $\delta\phi_{\rm MW}=$ 4, 6, 8 ${\rm rad~m^{-2}}$, respectively, with $F_{\rm BG}/F_{\rm MW} =$ 100. 
}
\label{f03}
\end{figure*}

We first consider full coverage of L band. This makes the situation simple and allows us to examine an importance of the P$_*$ band. Figure~\ref{f03} shows the results. The horizontal axis is the input $RM_{\rm IGMF}$ and the vertical axis is the center frequency of the P$_*$ band. The blue, green, and red lines show the contours on which $RM_{\rm IGMF}$ is determined with statistical errors of 30~\%, 20~\%, and 10~\%, respectively, based on the confidence intervals given by the Fisher information matrix.

The solid lines indicate the results with a 20~MHz bandwidth of P$_*$ band. For example, in figure~\ref{f03}(a), we safely obtain $RM_{\rm IGMF}=10$~${\rm rad~m^{-2}}$ with a statistical error less than 10~\%, if we choose the center frequency of P$_*$ band from 300~MHz to 750~MHz. We achieve a 10~\% error-level detection of $RM_{\rm IGMF}\sim 5$~${\rm rad~m^{-2}}$  (i.e. we can measure $RM_{\rm IGMF}=5\pm 0.5$~${\rm rad~m^{-2}}$), if we choose the center frequency of the P$_*$ band around 500~MHz.

The situation can be improved if a 40~MHz bandwidth of the P$_*$ band is available (dashed lines); in figure~\ref{f03}(a), an error level of 10~\% for $RM_{\rm IGMF}=5$~${\rm rad~m^{-2}}$ becomes a range from 370 MHz to 580 MHz. Even if we accept the error level of 30~\%, we can obtain $RM_{\rm IGMF}$ down to $\sim 1$~${\rm rad~m^{-2}}$ with a 40 MHz bandwidth at 450 MHz.

Figures~\ref{f03}(a)--(c) compare the effect of the intensity ratio, $f_{\rm BG}/f_{\rm MW}$. Surprisingly, we see that the intensity ratio does not significantly alter the results at least within the shown range, $f_{\rm BG}/f_{\rm MW}=$ 10--1000. Such independence on the intensity ratio can be seen in most of all our study in this paper. Therefore, hereafter we only show the results for the case of the intensity ratio $f_{\rm BG}/f_{\rm MW}=100$.

Figures~\ref{f03}(d)--(f) show the results for $\delta\phi_{\rm MW}=$ 4, 6, and 8 ${\rm rad~m^{-2}}$, respectively. Comparing with figure~\ref{f03}(b), the optimum frequency shifts to higher frequency. It indicates that the optimum frequency depends on the thickness of the Milky Way foreground emission. We also see that an increase of the bandwidth of P$_*$ band does not significantly improve the result, if $\delta\phi_{\rm MW}$ is relatively large.

%%%%%%%%%%%%%%%%%%%%%%%%%%%%%%%%%%%
\subsection{Impact of narrow L band}
\label{s3.2}

\begin{figure*}[t]
\begin{center}
\FigureFile(170mm,170mm){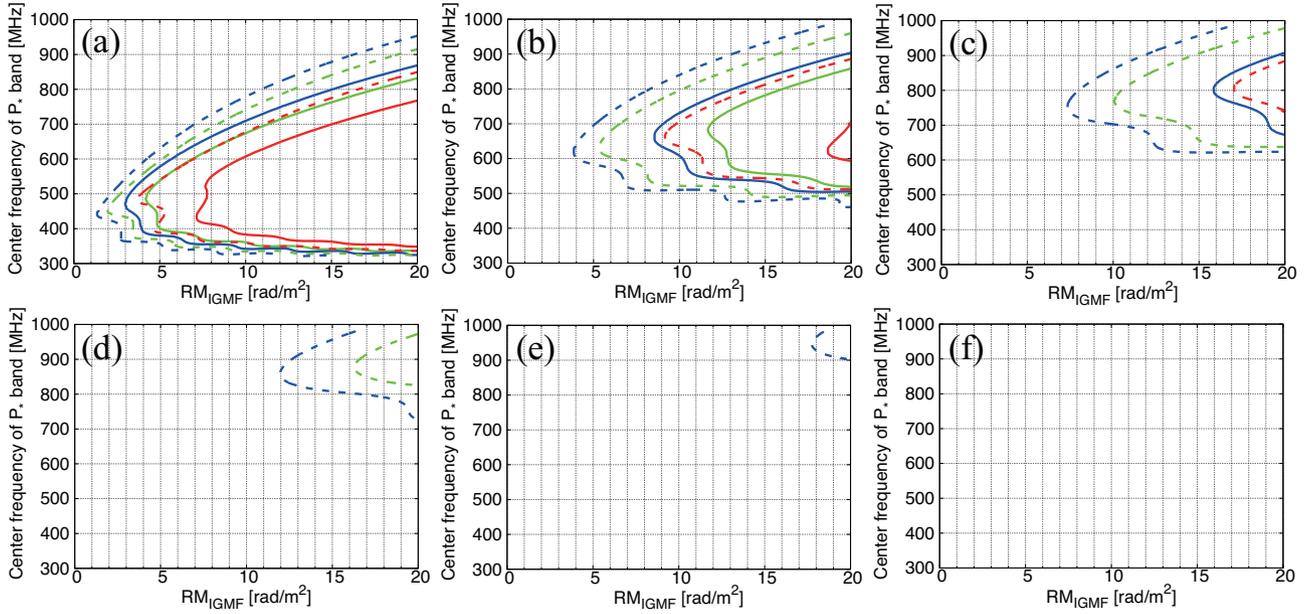}
\end{center}
\caption{
Same as figure~\ref{f03} but in the cases of the P$_*$ + L$_{\rm 14}$ + L$_{\rm 16}$ bands for $\delta\phi_{\rm MW}=$ 2, 4, 6, 8, 10, 12 ${\rm rad~m^{-2}}$ from (a) to (f), respectively, with $F_{\rm BG}/F_{\rm MW} =$ 100. 
}
\label{f04}
\end{figure*}

Figure~\ref{f04} show the optimum frequency of the P$_*$ band for the cases with the two narrow L bands (L$_{\rm 14}$ + L$_{\rm 16}$). Overall, the lack of data in L band makes the constraint of $RM_{\rm IGMF}$ worse, but we still obtain $RM_{\rm IGMF}$ with a reasonable error. For example, in figure~\ref{f04}(a), we obtain $RM_{\rm IGMF}=10$~${\rm rad~m^{-2}}$ with a statistical error less than 10~\%, if we choose the center frequency of P$_*$ band from 380~MHz to 610~MHz with a 20~MHz bandwidth. With a 40~MHz bandwidth, it improves to $RM_{\rm IGMF}=5$~${\rm rad~m^{-2}}$ for the center frequency of the P$_*$ band from 400~MHz to 550~MHz. 

Note that, compared to figure~\ref{f03}, we may need a more careful choice of the center frequency of the P$_*$ band; error levels quickly gets worse as the center frequency deviates from the optimum frequency. For example, in figure~\ref{f04}(a), when we set the center frequency at $\sim 500$~MHz, $\sim 650$~MHz, and $\sim 800$~MHz with a 40~MHz bandwidth, we achieve a 10~\% error-level detection of $RM_{\rm IGMF}=$ 4, 9, and 17~${\rm rad~m^{-2}}$, respectively.

We again see the dependence on $\delta\phi_{\rm MW}$. The optimum frequencies of the P$_*$ band are $\sim 500$~MHz, $\sim 650$~MHz, and $\sim 800$~MHz for $\delta\phi_{\rm MW}=$ 2, 4, and 6 ${\rm rad~m^{-2}}$, respectively. These optimum frequencies are similar to those seen in Fig~\ref{f03}. Contrary to the result seen in figure~\ref{f03}(d)--(f), increasing the bandwidth of the P$_*$ band substantially improves the results even for large $\delta\phi_{\rm MW}$ cases. If $\delta\phi_{\rm MW}$ exceeds $\sim 10$~${\rm rad~m^{-2}}$, however, the considering three narrowbands data can not constrain $RM_{\rm IGMF}$ within a reasonable error level.

%%%%%%%%%%%%%%%%%%%%%%%%%%%%%%%%%%%
\subsection{Improvement by narrow S band}
\label{s3.3}

\begin{figure*}[t]
\begin{center}
\FigureFile(170mm,170mm){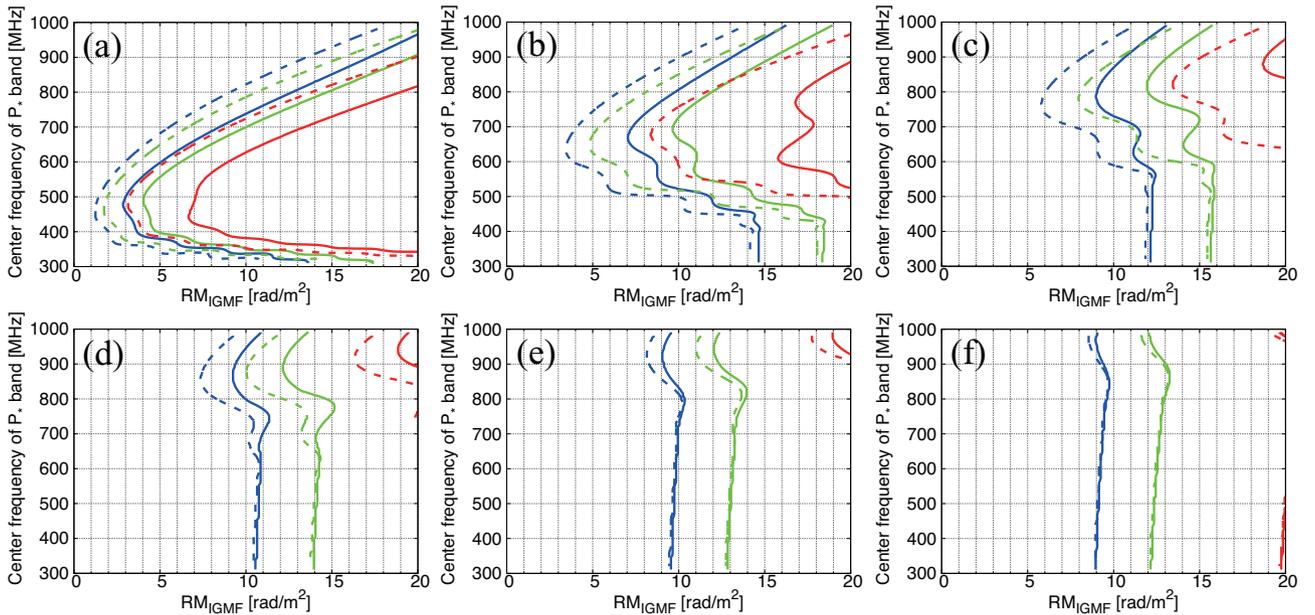}
\end{center}
\caption{
Same as figure~\ref{f03} but in the cases of the P$_*$ + L$_{\rm 14}$ + L$_{\rm 16}$ + S$_{\rm 27}$ bands for $\delta\phi_{\rm MW}=$ 2, 4, 6, 8, 10, 12 ${\rm rad~m^{-2}}$ from (a) to (f), respectively, with $F_{\rm BG}/F_{\rm MW} =$ 100.
}
\label{f05}
\end{figure*}

From the results in previous sections, we expect that high frequency (S band) data is useful when MW foreground is thicker (cases of low and mid galactic latitudes). Therefore, for the calculation in the previous section, we add the data of the S$_{\rm 27}$ narrowband and the results are shown in Fig.~\ref{f05}.

We see that the data at the S$_{\rm 27}$ band moderately improves the result for the cases of $\delta\phi_{\rm MW}=$ 2 and 4 ${\rm rad~m^{-2}}$. On the other hand, as expected, improvement is significant for the thicker cases. Increasing the bandwidth of P$_*$ band improves the results for thicker MW foreground. If we allow the error level of 30~\%, we obtain $RM_{\rm IGMF}$ down to $\sim 8$~${\rm rad~m^{-2}}$ with a 40 MHz bandwidth at 800--900 MHz, for the cases with $\delta\phi_{\rm MW}\sim 8$--10~${\rm rad~m^{-2}}$.

%%%%%%%%%%%%%%%%%%%%%%%%%%%%%%%%%%%
%%%%%%%%%%%%%%%%%%%%%%%%%%%%%%%%%%%
\section{Discussion and Summary}
\label{s4}

We found that the optimum frequency depends on the thickness of the Faraday spectrum for the foreground Milky Way emission. The optimum frequencies are $\sim 500$~MHz, $\sim 650$~MHz, $\sim 800$~MHz, $\sim 950$~MHz for $\delta\phi_{\rm MW}=$ 2, 4, 6, 8 ${\rm rad~m^{-2}}$, respectively, and it reaches $\sim 1400$~MHz if $\delta\phi_{\rm MW} = 15$~${\rm rad~m^{-2}}$. We find that the optimum frequency is close to the frequency at which foreground Milky Way emission is significantly depolarized at the observer frame (figure~\ref{f02}). Such depolarization is seen in the polarized-intensity spectrum, for example, at $\lambda^2 \sim 0.2$~${\rm m}^2$ in the bottom panel of figure~\ref{f02}. This depolarization is classified into differential Faraday rotation depolarization \citep{sok98, ars11}. \citet{ars11} investigated the optimum wavelength ($\lambda_{\rm opt}$) of the maximum polarized emission according to differential Faraday rotation, and proposed the equation of the optimum wavelength as
\begin{equation}\label{eq06}
|\sin k | - k |\cos k| = 0,
\end{equation}
where we consider a flat spectral index (the case of $\alpha=0$ in \cite{ars11}) and $k=2 |RM| \lambda_{\rm opt}^2$. The optimum frequencies that we found is in broadly agreement with the solution of $k= 2.0288~({\rm radian})$ for an effective RM value of $|RM|\sim 1.3 \delta \phi_{\rm MW}$. Therefore, the optimum frequencies can be explained by the depolarization theory.

The above depolarization frequency, i.e. the optimum frequency, depends on the model Faraday spectrum of the Milky Way; we have considered a Gaussian shape and the intrinsic polarization angle is constant. We can consider more complicated, realistic FDFs of polarized sources \citep{ide14b}. However, this work focuses on a typical, global solution of the optimum frequency. A specific model is beyond the scope of this work and it will be considered in a separate paper. Nevertheless, if an actual Faraday spectrum of the Milky Way deviates from the Gaussian, the depolarization frequency can change. Note that the constraint on $RM_{\rm IGMF}$, i.e. the gap between MW and BG, primarily depends on the edge of the Faraday spectrum of the Milky Way rather than a detailed profile of the Faraday spectrum of the Milky Way.

Throughout this paper, the total intensity is independent on the frequency so that a flat spectral index is considered. If we consider a steep spectrum, the intensity of the P$_*$ band becomes brighter and the signal to noise ratio becomes better by several times. This may result in better constraint on $RM_{\rm IGMF}$, because we obtain better quality of data at the P$_*$ band. We will address this effect more quantitatively in future, since Faraday tomography considering a non-zero spectral index is under development.

The intensity ratio between the background and foreground sources does not significantly change the results, and exceptionally bright (Jy-level) background sources are available to this work of exploring the IGMF. Therefore, our method can be applicable for background, linearly-polarized FRBs. Meanwhile, detection of the Milky Way foreground would be more challenging. We have considered the signal-to-noise ratio of 10 for MW in each 1 MHz channel. If the noise level is higher, it seriously impacts on the detection of the Milky Way. Moreover, an interferometric observation may suffer from the missing flux of diffuse foreground emission. 

Although we introduced RFIs in Kashima as an example, our results do not depend on where and how the polarized intensity spectrum is obtained. Therefore, our results can be applicable to other current radio facilities and even the future telescopes such as the SKA.

A possible recipe to confront this Milky Way foreground issue would be that we combine another single-dish observation of diffuse Milky Way foreground. Comparison between on-source and off-source observations is also useful, where the off-source observation measures a nearby sky sharing almost the same foreground. These follow-up observations confirm the diffuse foreground and decide the edge of the Faraday spectrum of the Milky Way. If we have two background sources located closely each other, we can apply another methodology, case (ii), discussed in \citet{aka14a}.

In summary, we studied optimum frequencies to constrain Faraday rotation measure (RM) due to the IGMF by means of Faraday tomography. The frequency resolution of 1 MHz has been considered throughout this work. Using a simple model and Fisher information matrix, we find that multiple narrowband data in the UHF provides a reasonable constraint on the RM due to the IGMF. With data at 1400 MHz and 1600 MHz, RM$_{\rm IGMF} \sim 10$~${\rm rad~m^{-2}}$ toward a high Galactic latitude is detectable with less than 10~\% error, if we choose the center frequency of the P$_*$ band around 400 -- 700 MHz with a 40 MHz bandwidth.

\vskip 12pt

This work was supported in part by JSPS KAKENHI Grant Numbers JP15H05896 (KT), JP16H05999 (KT), JP16K13788 (T. Aoki), JP17K01110 (T. Akahori, KT), and Bilateral Joint Research Projects of JSPS (KT). Numerical computations were carried out on PC cluster at Center for Computational Astrophysics, National Astronomical Observatory of Japan.

\appendix 
%%%%%%%%%%%%%%%%%%%%%%%%%%%%%%%%%%%%%%%%%%%
%%%%%%%%%%%%%%%%%%%%%%%%%%%%%%%%%%%%%%%%%%%
\section*{A. RFI Observation at Kashima}

We investigated radio environment at NICT Kashima in daytime and nighttime in August 28, 2017, JST. It was cloudy and slightly windy. We measured radio spectra between 100 MHz to 1000 MHz using a discorne antenna and between 1000 MHz to 3000 MHz with a tear-drop antenna. Antennas were placed at a pedestal of the rooftop of the operation center so as to ensure that a height of the antenna exceeds that of the metal fence enclosing the rooftop. Band characteristics were measured by replacing an antenna into a terminator and were removed from the RFI data. The data was visualized with a spectrum analyzer. Modes of the spectrum analyzer were instantaneous (blue) and 5 minutes max-hold (orange), where we set the resolution bandwidth (RBW) 1MHz and the video bandwidth (VBW) 1kHz. The results obtained around 11 PM is shown in figure.~\ref{f01}. We observed that the shown spectrum is time-dependent.

The beam patterns of the both antennas are torus-like and the most sensitive to the ground, horizontal direction. They have only capabilities to capture vertical polarization with respect to the ground plane. Therefore, the results do not fully cover RFIs from directly above and RFIs of horizontal polarization from all horizontal directions. For instance, most of television broadcasts in Japan are horizontal polarization, which is not sensitive in our experience. Therefore, its effect is likely underestimated. Since the antenna is omnidirectional, it is difficult to identify the locations of the origins of the RFIs.

\end{document}